\RequirePackage{lineno}
\documentclass[article,prl,
twocolumn,
amsmath,amssymb,
amsmath,amssymb, aps,
]{revtex4}

\usepackage[dvipsnames]{xcolor}
\usepackage{epsfig}
\usepackage{amsfonts}
\usepackage{amsmath}
\usepackage{wasysym}
\usepackage{amssymb}
\usepackage{bm}
\usepackage{float}
\usepackage{enumitem}

\usepackage{pdflscape}
\usepackage{graphicx}
\usepackage{ulem}

\usepackage[utf8]{inputenc}
\usepackage{bbm} 				
\usepackage{epstopdf}				
\usepackage{verbatim}    			
\usepackage{sidecap}                
\usepackage{indentfirst}

\newcommand{\be}{\begin{equation}}
\newcommand{\ee}{\end{equation}}
\newcommand{\bea}{\begin{eqnarray}}
\newcommand{\eea}{\end{eqnarray}}
\newcommand{\bfr}{\mathbf{r}}

\definecolor{armygreen}{rgb}{0.0, 0.5, 0.0}

\begin{document}

\title{Stochastic Model of Organizational State Transitions in a Turbulent Pipe Flow}

\author{Robert Jäckel$^{1,3}$, Bruno Magacho$^{2,3}$, Bayode Owolabi$^{2,3}$, \\
Luca Moriconi$^{2,3}$\footnote{Corresponding author: moriconi@if.ufrj.br}, David J.C. Dennis$^{3}$, and Juliana B.R. Loureiro$^{1,3}$}
\affiliation{$^1$Programa de Engenharia Mecânica, Coordenação dos Programas de Pós-Graduação em Engenharia,
Universidade Federal do Rio de Janeiro, C.P. 68503, CEP: 21941-972, Rio de Janeiro, RJ, Brazil}
\affiliation{$^2$Instituto de F\'\i sica, Universidade Federal do Rio de Janeiro,
Av. Athos da Silveira Ramos 149, CEP: 21941-909, Rio de Janeiro, RJ, Brazil
}
\affiliation{$^3$Interdisciplinary Center for Fluid Dynamics,
Universidade Federal do Rio de Janeiro, R. Moniz Arag\~{a}o 360, CEP: 21941-594, Rio de Janeiro, Brazil}


\begin{abstract}
Turbulent pipe flows exhibit organizational states (OSs) that are labelled by discrete azimuthal wavenumber modes and are reminiscent of the traveling wave solutions of low Reynolds number regimes. The discretized time evolution of the OSs, obtained through stereoscopic particle image velocimetry, 
is shown to be non-Markovian for data acquisition carried out at a structure-resolved sampling rate. In particular, properly defined time-correlation functions for the OS transitions are observed to decay as intriguing power laws, up to a large-eddy time horizon, beyond which they decorrelate at much faster rates. We are able to establish, relying upon a probabilistic description of the creation and annihilation of streamwise streaks, a lower-level {\it{Markovian}} model for the OS transitions, which reproduces their time-correlated behavior with meaningful accuracy. These findings indicate that the OSs are distributed along the pipe as statistically correlated packets of quasi-streamwise vortical structures.
\end{abstract}


\maketitle
Notwithstanding the large body of knowledge accumulated since the landmark experiments of Reynolds \cite{reynolds1}, turbulent pipes comprise flow patterns which have remained surprisingly unsuspected until recent years. They can be depicted as relatively organized sets of wall-attached low-speed streaks coupled to pairs of counter-rotating quasi-streamwise vortices \cite{hof_etal,schneider_etal,dennis_sogaro}. These {\it{organizational states}} (OSs) actually characterize the turbulent velocity fluctuations at high Reynolds numbers and are topologically similar to traveling waves -- a class of exact (but unstable) low-Reynolds number solutions of the Navier-Stokes equations \cite{faisst_eck,wedin_kers}. 

As for traveling waves, the OSs can be classified by the number of low-speed streaks they contain. Observation tells us, however, that this quantity changes in an apparently random way along the turbulent pipe. For the sake of illustration, Fig.~1 shows a transition between OSs, visualized from a pair of cross-sectional snapshots of the flow obtained through stereoscopic particle image velocimetry (sPIV).

The existence of spatial transitions among the OS modes suggests, within the perspective of dynamical systems, that the turbulent pipe flow could be described as a chaotic attractor and its unstable periodic orbits in a phase space of much reduced dimensionality \cite{gibson_etal,moehlis_etal,budanur_etal, yalniz_etal,marensi_etal}. In connection with this circle of ideas, we are motivated to study the OS transitions in the framework of stochastic processes, focusing particular attention on their recurrent dynamics.

\begin{figure}[ht]
\includegraphics[width=0.47\textwidth]{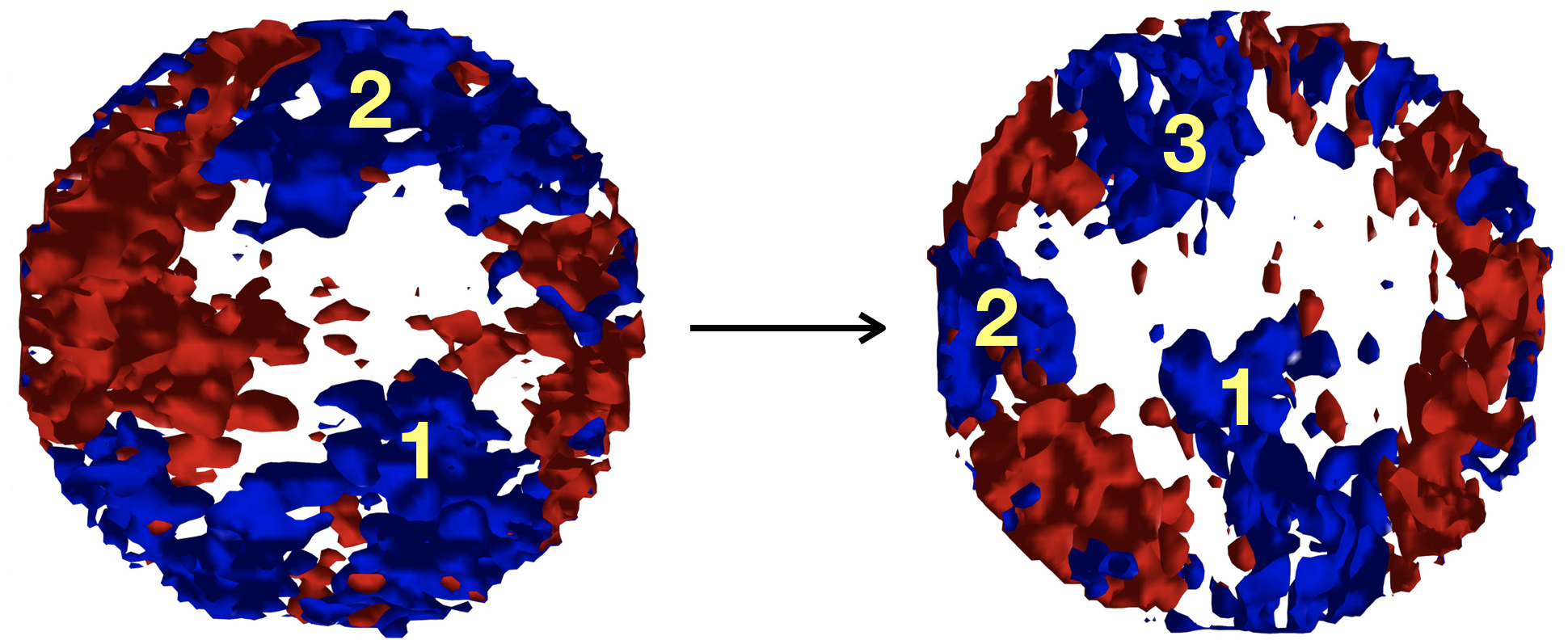}
\caption{Example of a transition between organized states, as sampled out from our measurements, which are associated to two and three low-speed streaks. Blue and red colors refer, respectively, to negative and positive streamwise velocity fluctuations around the mean (the systematic procedure to ascertain a well-defined number of low-speed structures to a given flow snapshot is discussed in the text).}
\end{figure}

To start, let {\hbox{$u=u(r,\theta)$}} be, in polar coordinates, the fluctuating streamwise component of the velocity field defined over a fixed pipe's cross-sectional plane. We may introduce, accordingly, the instantaneous spectral power density,
\be
I(k_n) =  \left |\int_0^{2 \pi} d \theta e^{ i k_n  \theta  }  
f_{uu}(r_0,\theta)    \right |^2 \ , \ 
\ee
where
\be 
f_{uu}(r_0,\theta) =
\int_0^{2 \pi} d \theta' 
u(r_0, \theta') u(r_0,\theta' + \theta)  \ , \ 
\ee
$k_n = n \in \mathbb{Z}^+$ is an azimuthal wavenumber, and $r_0$ is a reference radial distance which falls within the  log-region of the pipe's turbulent boundary layer. Empirical evidence shows that $I(k_n)$ is in general peaked at some clearly dominant wavenumber $\bar k$ (to be identified to the number of snapshotted low-speed streaks), which can be used to label the probed velocity profile $u(r,\theta)$. As time evolves, $I(k_n)$ changes, and so does the wavenumber position of its dominant peak.
Therefore, if $u(r,\theta)$ is recorded at equally spaced time intervals $\Delta$, the dynamical evolution of the pipe turbulent field can be mapped into the stochastic process
\be 
{\cal{S}} \equiv \{ \bar k(t), \bar k(t + \Delta), \bar k(t + 2\Delta), \ ... \ \} \ . \ \label{S_process}
\ee
In order to investigate the still very open statistical properties of ${\cal{S}}$, we have performed a pipe flow experiment, at Reynolds number Re = 24415, in the large pipe rig facility of the Interdisciplinary Nucleus for Fluid Dynamics (NIDF) at the Federal University of Rio de Janeiro. The pipe's diameter and length are, respectively, $D = 15$ cm and {\hbox{$L = 12$ m}}. By means of sPIV, with sampling rate of {\hbox{$10$ Hz}} (i.e., $\Delta = 0.1$ s), we have collected $10^4$ cross-sectional snapshots of the flow, each one containing the three components of the turbulent velocity field over a uniform grid of size $78 \times 78$. It turns out that all the observed OS modes fall into the range $0 \leq \bar k \leq \bar k_{max} = 10$.

Our experimental data has been validated with the help of previous benchmark pipe flow experiments \cite{toonder_nieu}, through the inspection of the performance of first and second order single-point statistics for the streamwise component of velocity field. We have also attained a further validation of the entire measured velocity field, from the evaluation of particularly defined streamwise velocity-velocity correlation functions conditioned on the OS modes $\bar k$, more precisely,
\begin{equation}
R_{uu}(\Delta \bfr | \bar k) \equiv \mathbb{E} [ u(\bfr_0) u(\bfr_0 + \Delta \bfr) | \bar k ] \ , \ 
\end{equation}
which has its level curves depicted in Fig.~2, for the case $\bar k =5$, in close correspondence with the results of {\hbox{Ref. \cite{dennis_sogaro}}.

\begin{figure}[t]
\includegraphics[width=0.47\textwidth]{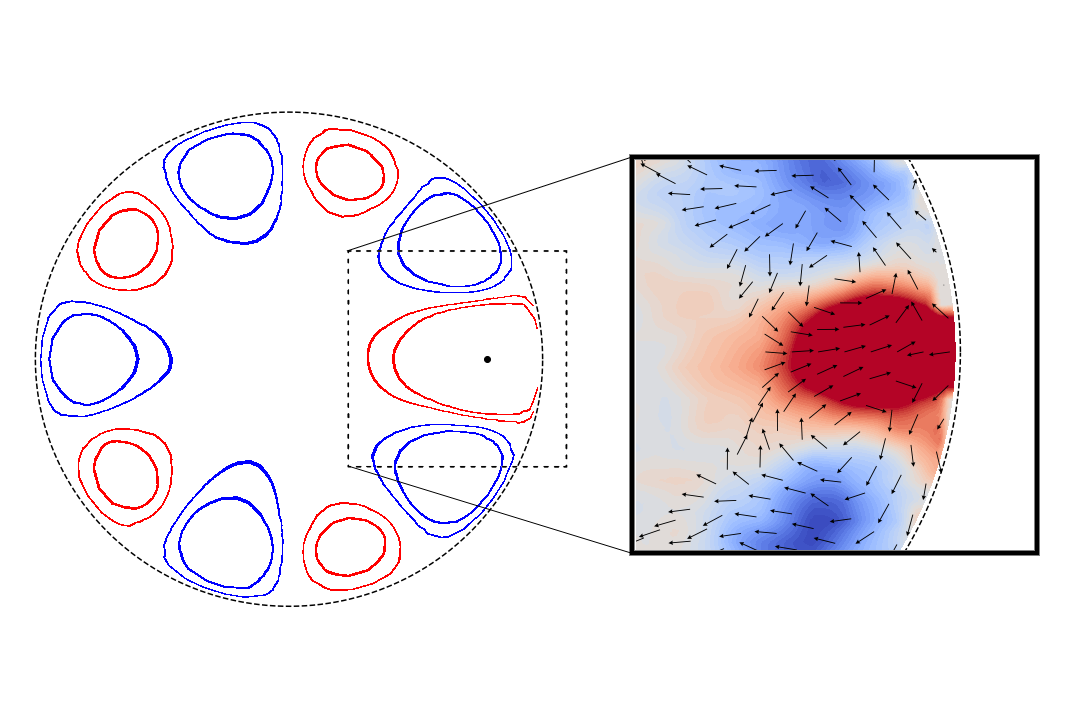}
\vspace{-1.0cm}
\caption{Statistical results for the OS mode $\bar k = 5$. Left image: positive (red) and negative (blue) level curves of $R_{uu}$, defined by $| R_{uu}({\bf{r}}-{{\bf{r}}}_0 | \bar k)| =$ 5$\%$ and 10$\%$ of $(R_{uu})_{max}$, with the reference point ${\bf{r}}_0$ depicted as a black dot. Right image: a closer look at the averaged streamwise velocity fluctuations (red for positive, blue for negative), conditioned on $u({\bf{r}}_0) >0$. The cross-sectional averaged velocity field reveals the vortical structures that are usually coupled with velocity streaks.}
\end{figure}
The first immediate question that can be raised about the stochastic process $\cal{S}$ is whether it is Markovian or not. Of course, while it is not possible to answer this in full rigor, one may check if the Chapman-Kolmogorov (CK) equation holds for the time series (\ref{S_process}), a necessary condition for $\cal{S}$ to be Markovian \cite{erhan}. The CK equation would imply that the eigenvalues of the transition probability matrix for OS modes separated by the time interval $h \Delta$ can be represented, in some arbitrary ordering, as the set of powers $\{ \lambda_1^h, \lambda_2^h, ..., \lambda_{\bar k_{max}}^h \}$. A straightforward computation of the transition matrix eigenvalues for the cases $h=1$ and $h=2$ indicates, however, that $\cal{S}$ is not Markovian; see Fig.~3. 

We expect that the decimated process for $h$ large enough is essentially Markovian, since in this situation the OS modes become weakly correlated. The transition to Markovian behavior can be alternatively addressed from the analysis of correlation functions which we introduce as it follows.
Taking $0 \leq m, m' \leq \bar k_{max}$, let $V_m(t)$ and $M_{m'm}(t)$ be, respectively, the components of vector and matrix valued stochastic processes derived from $\cal{S}$ as 
\begin{equation}
 V_{m}(t) =
    \begin{cases}
      1, & \text{if~~$\bar k(t) = m$} \\
      0, & \text{otherwise}
    \end{cases}       
\end{equation}
and
\begin{equation}
M_{m'm}(t) = 
    \begin{cases}
      1, & \text{if~~$\bar k(t) = m$~~and~~$\bar k(t+\Delta) = m'$} \\
      0, & \text{otherwise~.}
    \end{cases}       
\end{equation}
\begin{figure}[t]
\includegraphics[width=0.47\textwidth]{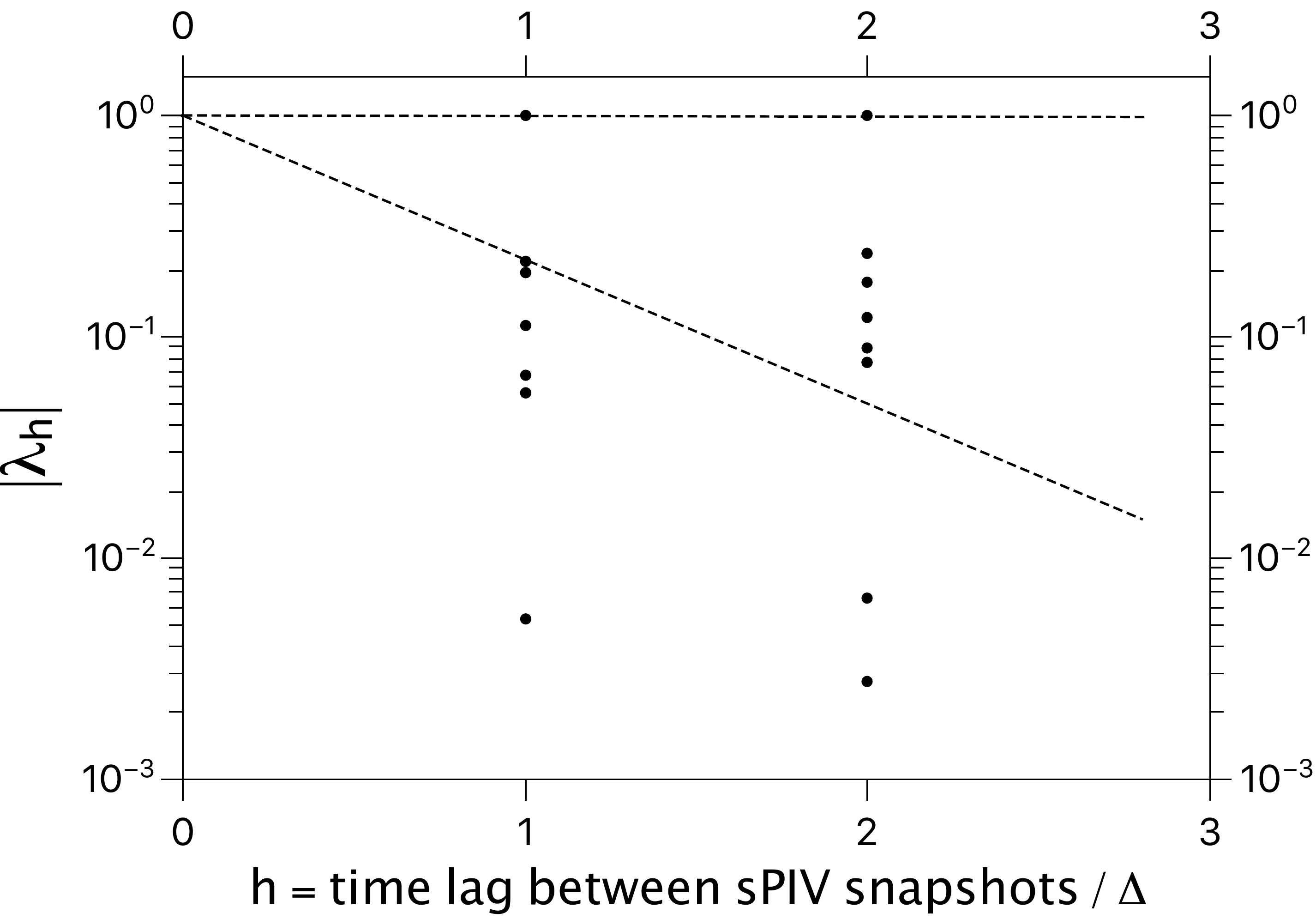}
\caption{Eigenvalues of the probability transition matrices for the original process ($h=1$) and a decimated one ($h=2$). The dashed lines should intercept eigenvalue pairs if $\cal{S}$ were a Markovian process.}
\end{figure}
\begin{figure}[t]
\includegraphics[width=0.47\textwidth]{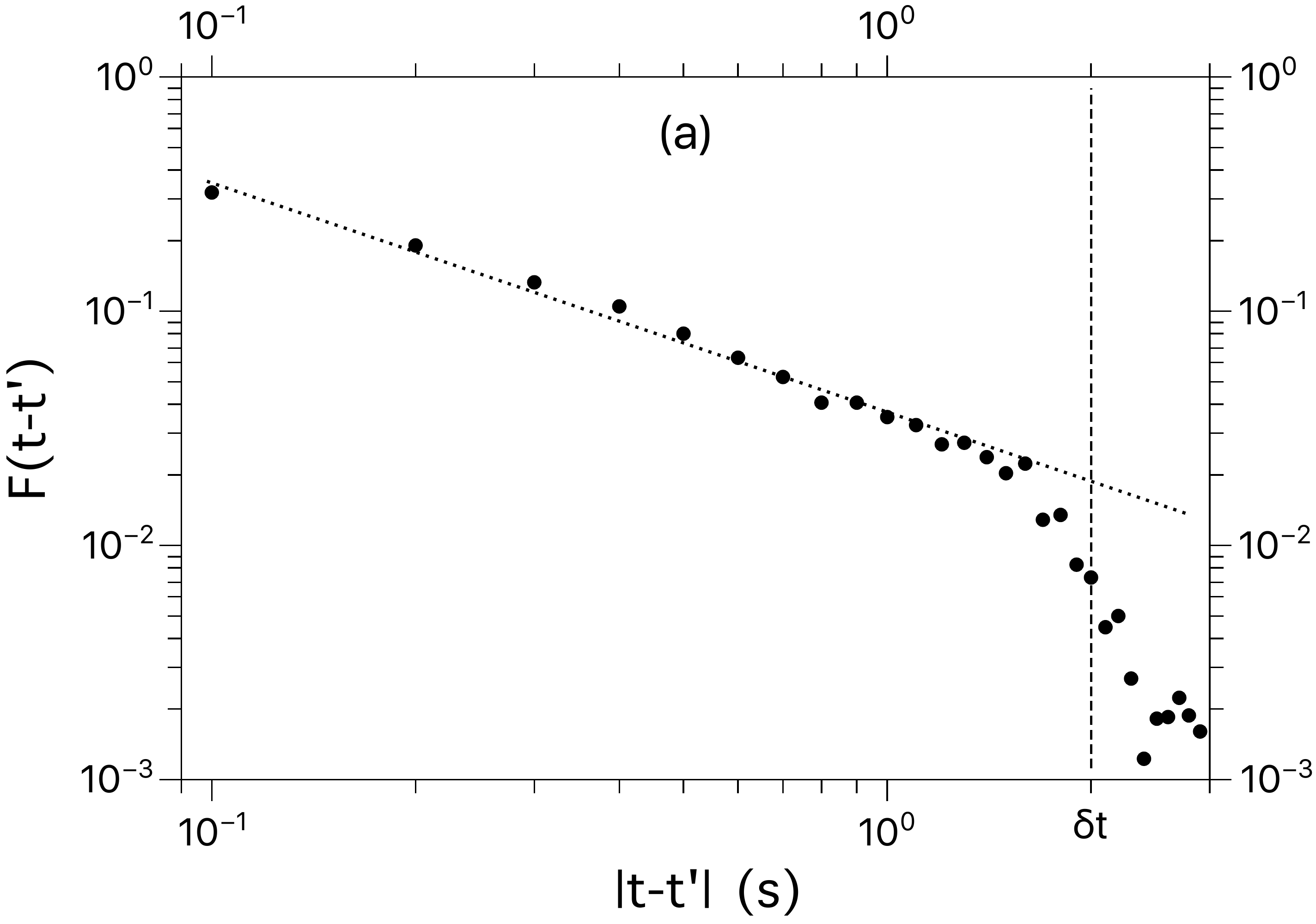}
\includegraphics[width=0.47\textwidth]{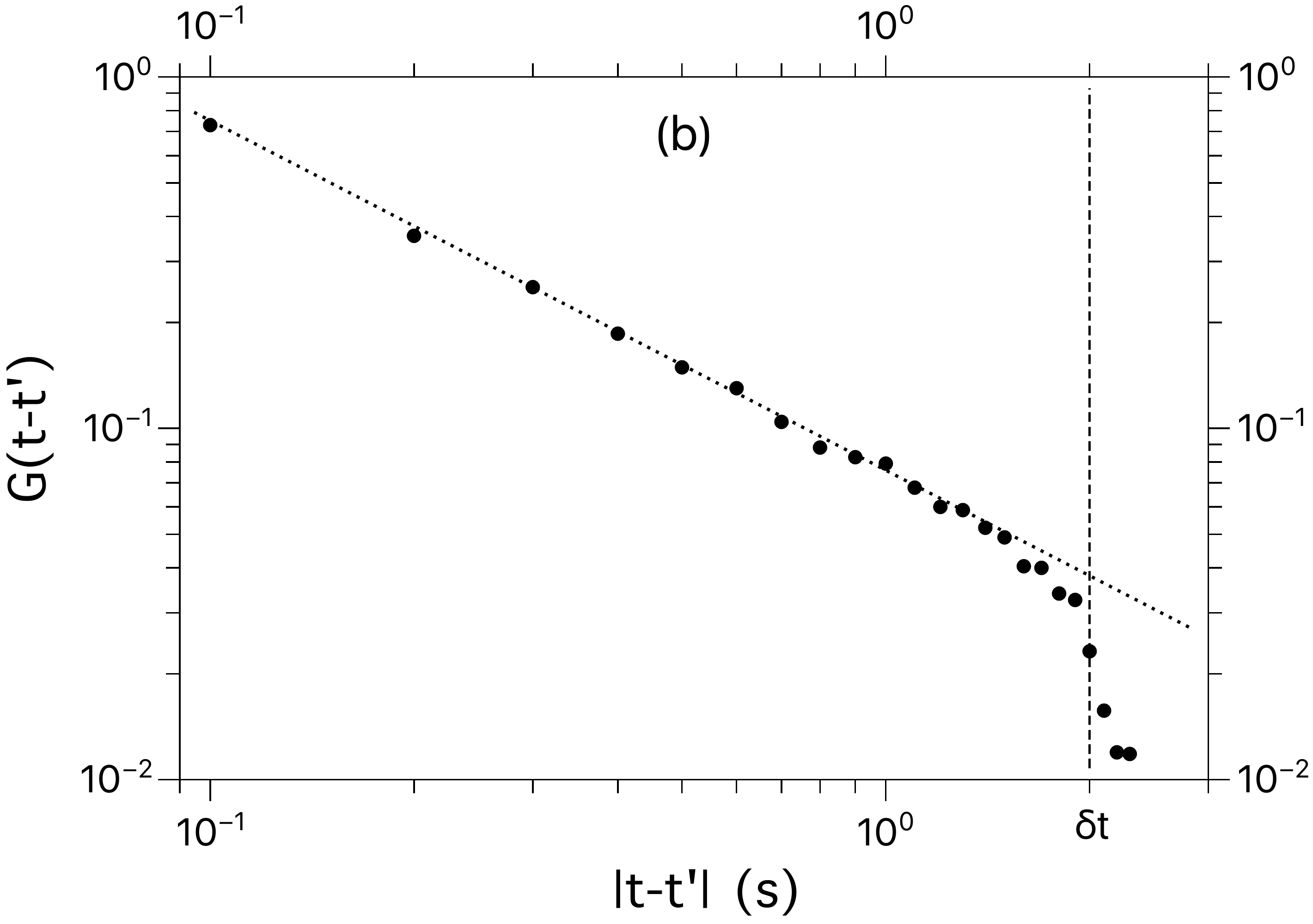}
\caption{The time-dependent correlation functions defined in (\ref{GF}) are noticed to decay as power laws for {\hbox{$|t - t'| \leq  \delta t \approx 2$ s}}. The dotted lines in (a) and (b) have scaling exponent $-1$ for both $F(t-t')$ and $G(t-t')$.}
\end{figure}
Define, now, the correlation functions
\begin{eqnarray}
\tilde F(t-t') &\equiv& \mathbb{E}[{\bf{V}}(t) \cdot {\bf{V}}(t')] - 
(\mathbb{E}[{\bf{V}}])^2 \ , \ {\hbox{~~~~~~~~~~~~~~~~}}   \\
\tilde G(t-t') &\equiv& {\hbox{Tr}} \left \{ \mathbb{E}[{\bf{M}^T}(t){\bf{M}}(t')] - 
\mathbb{E}[{\bf{M}}]{\bf{^T}} \mathbb{E}[{\bf{M}}] \right \} \ , \ 
\end{eqnarray}
and their normalized versions,
\be
F(t-t') \equiv \frac{\tilde F(t-t')}{\tilde F(0)} \ , \
G(t-t') \equiv \frac{\tilde G(t-t')}{\tilde G(0)} \ . \ \label{GF}
\ee 
It is not difficult to see that $F(t-t')$ and $G(t-t')$ describe, respectively, the correlations of {\it{returning}} OS modes and transitions which are apart from each other by the time interval $|t-t'|$.
They are plotted in Fig.~4 and are noticed to have interesting power law decays (with the same approximate scaling exponent
$-1$) up to $|t-t'| \equiv \delta t \approx 20 \Delta = 2$ s, which suggests some sort of self-similarity across the spatial distribution of about ten OS modes (their mean lifetime is 
{\hbox{$0.2$ s $\approx \delta t/10$}}). For time separations larger than $\delta t$, the correlation functions become suddenly undersampled, meaning that they crossover to a faster law of decay, probably exponential, as it should be for the putative asymptotic Markovian behavior of a large-time decimated $\cal{S}$.

It is worth emphasizing that the non-Markovian nature of $\cal{S}$ does not mean at all that it cannot be modeled as a Markov process defined in terms of lower-level state variables. In this connection, it is reasonable to assume that there is a combinatoric degeneracy factor 
\be
\Omega(\bar k_{max},m) = \binom{\bar k_{max}}{m}
\ee
\\
\noindent associated to a given OS mode $\bar k = m$.  We simply mean here that the $m$ wall-attached low-speed streaks can be spatially arranged for this particular mode in $\Omega(\bar k_{max},m)$ different ways, since the pipe's cross-sectional plane is taken to hold at most $\bar k_{max}$ low-speed {\it{streak channels}}. The phase space of the ``microscopic" state variables for the underlying Markovian model of $\cal{S}$ is spanned, therefore, by all the possible sets of $\bar k_{max}$ {\it{streak bits}}, {\hbox{$X \equiv \{ s_1, s_2, ..., s_{\bar k_{max}} \}$}}, where
\begin{equation}
s_i =
    \begin{cases}
      1, & \text{if the $i$-th streak channel is active} \\
      0, & \text{otherwise~.}
    \end{cases}       
\end{equation}
We postulate, now, that the time evolution of the microscopic states $X$ is produced from the independent fluctuations of streak bits, which have persistence probabilities that depend on the total number of active streak channels, that is the OS label $m$. In this way, we define $q_m$ and $p_m$ to be the persistence probabilities for any given streak bit to keep its value 0 or 1, respectively, along subsequent sPIV snapshots. There are, thus, four different types of streak bit flips, which appear in different occurrence numbers for a given OS mode transition, as summarized in Table~I.
\begin{table}[h]
\centering
  \begin{tabular}{ | c | c | c | }
    \hline
    Transition Type & \# of Streak Channels & Transition Prob.\\ \hline
     0 $\rightarrow$ 0 & $n_1$ & $q_m$ \\ \hline
         0 $\rightarrow$ 1 & $n_2$ & $1-q_m$ \\ \hline
             1 $\rightarrow$ 0 & $n_3$  & $1-p_m$\\ \hline
     1 $\rightarrow$ 1 & $n_4$  & $p_m$ \\
    \hline
  \end{tabular}
\caption{Definition of the four possible transition types for the streak channel states, together with the notations for their occurrence numbers and individual transition probabilities. Above, $m = n_3 + n_4$ labels the OS mode.}
\label{tab}
\end{table}

The parameters reported in Table I are related to the OS mode transition $m \rightarrow m'$, where $m = n_3 + n_4$ and $m' = n_2 +n_4$. The transition probability between any specific pair of associated microstates is, as a consequence, $q_m^{n_1}(1-q_m)^{n_2} (1-p_m)^{n_3} p_m^{n_4}$. Taking into account, furthermore, the role of degeneracy factors, we may write the transition probability between the OS modes $m$ and $m'$ as
\begin{widetext}
\begin{eqnarray}
T_{m'm} &=&  \binom{\bar k_{max}}{m}^{-1} \sum_{n_1=0}^{\bar k_{max}}
\sum_{n_2=0}^{\bar k_{max}} \sum_{n_3=0}^{\bar k_{max}} \sum_{n_4=0}^{\bar k_{max}} 
\delta(n_1+n_2+n_3+n_4,{\bar k_{max}}) 
\delta(n_3+n_4,m) \delta(n_2+n_4,m') \times \nonumber \\
&\times& \binom{{\bar k_{max}}}{n_1} \binom{{\bar k_{max}}-n_1}{n_2} \binom{{\bar k_{max}}-n_1-n_2}{n_3} q_m^{n_1}(1-q_m)^{n_2} (1-p_m)^{n_3} p_m^{n_4}
\ . \  \label{Tmatrix}
\end{eqnarray}
\end{widetext}
Using, from now on, $\bar k_{max} = 10$, the Markovian model just introduced may not appear very phenomenologically attractive at first glance, since $T_{m'm}$ is parametrized by a large number of unknown parameters ($q_0,q_1, ...,q_9$ and $p_1,p_2, ...,p_{10}$). Note, however, that there are, in principle, 90 independent entries in the empirical transition matrix (the one derived from the sPIV measurements), so the model is rather underdetermined (as we would expect for a phase-space reduced description of turbulent fluctuations).
\begin{figure}[b]
\includegraphics[width=0.47\textwidth]{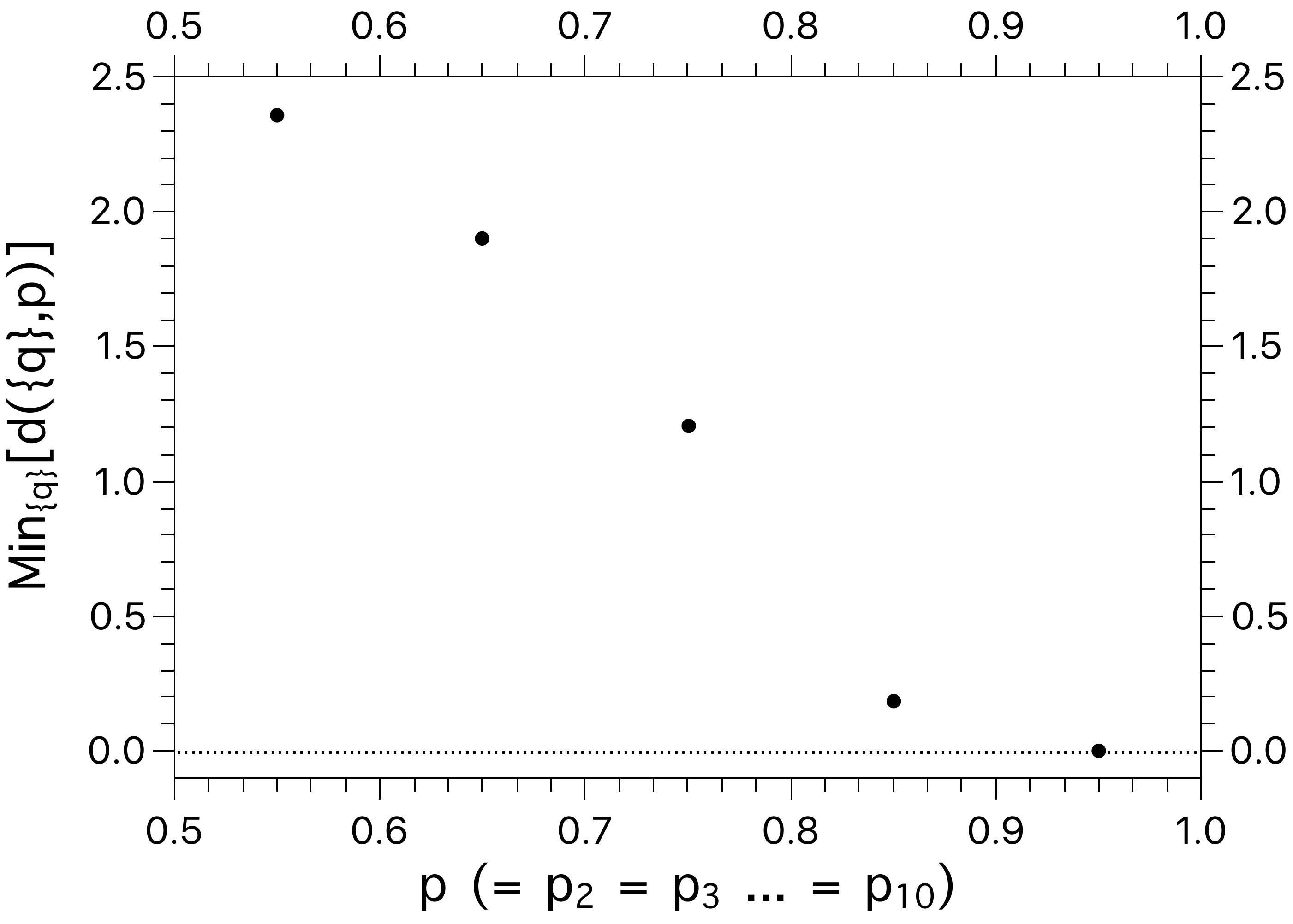}
\caption{Minimization of the quadratic distance $d(\{ q \},p)$ for various 
values of $p$.}
%
\includegraphics[width=0.47\textwidth]{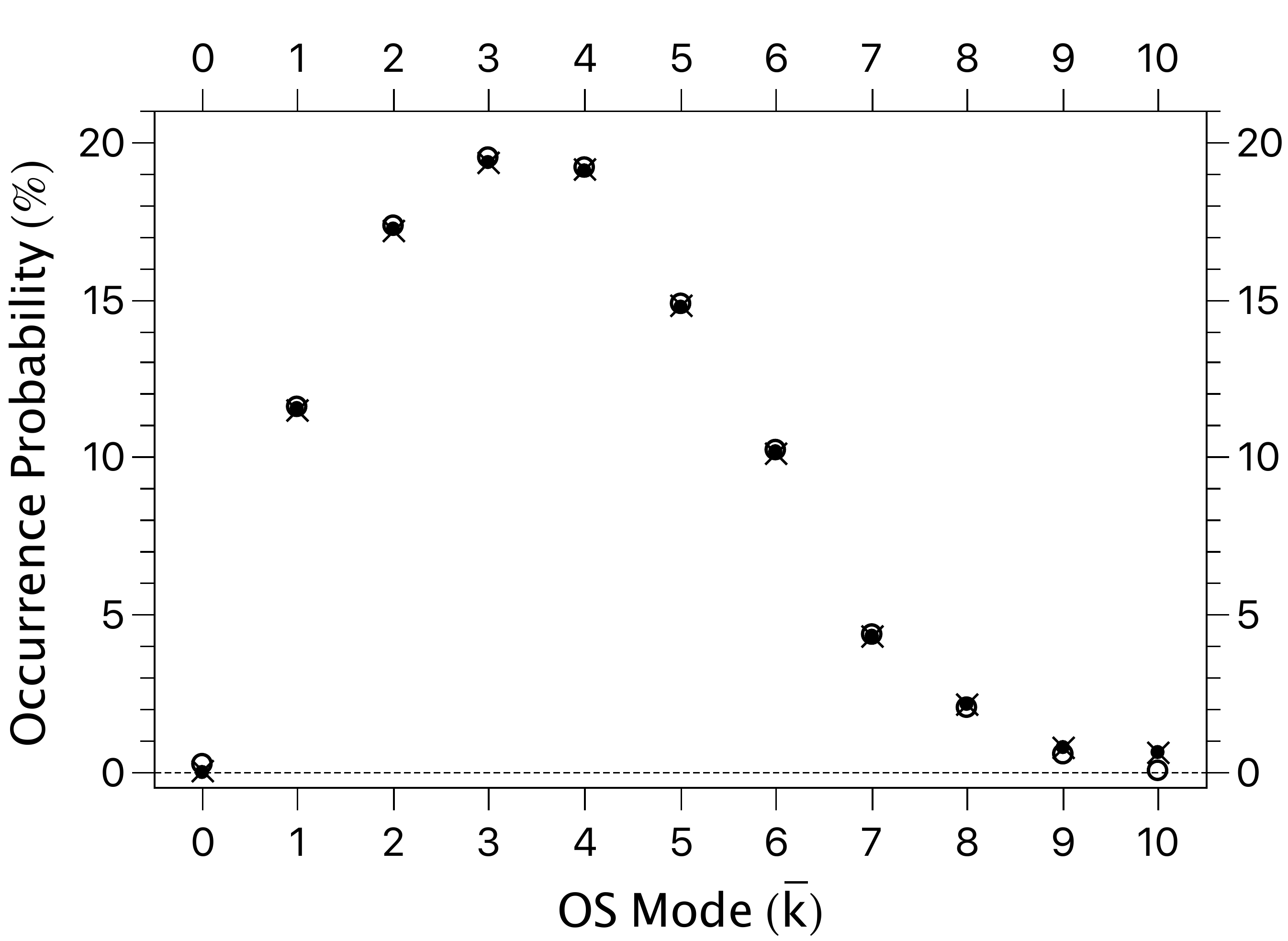}
\caption{The occurrence probability of OS modes obtained from the experiment (dots) and from the stochastic model (open circles: $p=0.86$; crosses: $p=0.95$), defined by the transition matrix elements (\ref{Tmatrix}).}
\end{figure}

Instead of attempting to provide a detailed and computationally costly model of the empirical transition matrix, we address a much simpler approach, where we focus on the asymptotic probability eigenvector of the modeled transition matrix,
\be 
{\mathbb{P}} = (P_1, P_2, ...,P_{10}) \ , \ \label{aprob}
\ee
which satifies to $\mathbb{TP = P}$, that is,
$
\sum_{m=0}^{10} T_{m'm} P_m = P_{m'} 
$.
Here, $P_m$ is the probability that the OS mode $m$ be observed in the statistically stationary regime. In an analogous way, denoting by ${\mathbb{P}_\infty}$ the empirical probability vector, determined 
from the sPIV measurements, we are interested to find the set of probabilities $q_m$ and $p_m$ that minimize the quadratic error
\be 
d(\{q_m\},\{p_m\})\equiv  ||{\mathbb{P}} - {\mathbb{P}_\infty} ||^2 \ . \ \label{qerror}
\ee
While, as already commented, the original problem is underdetermined, the optimization scheme related to 
Eq.~(\ref{qerror}) is not: as a matter of fact, we would have to model the 9 independent probability entries of (\ref{aprob}) by means of the 20 probability parameters $q_m$ and $p_m$. To reduce this large overdeterminacy, we rely on a few phenomenological inputs:
\vspace{0.2cm}

\begin{figure}[t]
\includegraphics[width=0.47\textwidth]{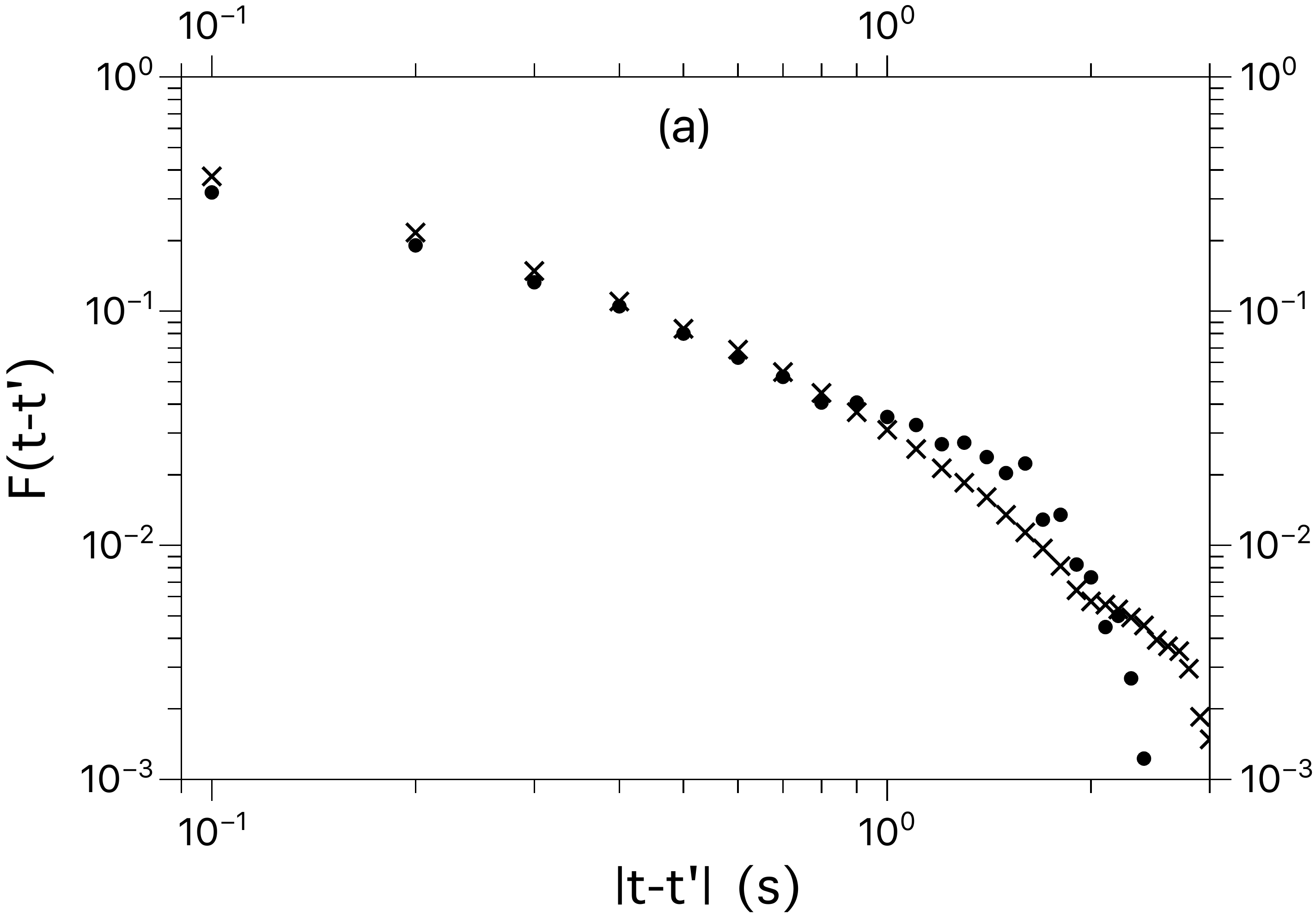}
\includegraphics[width=0.47\textwidth]{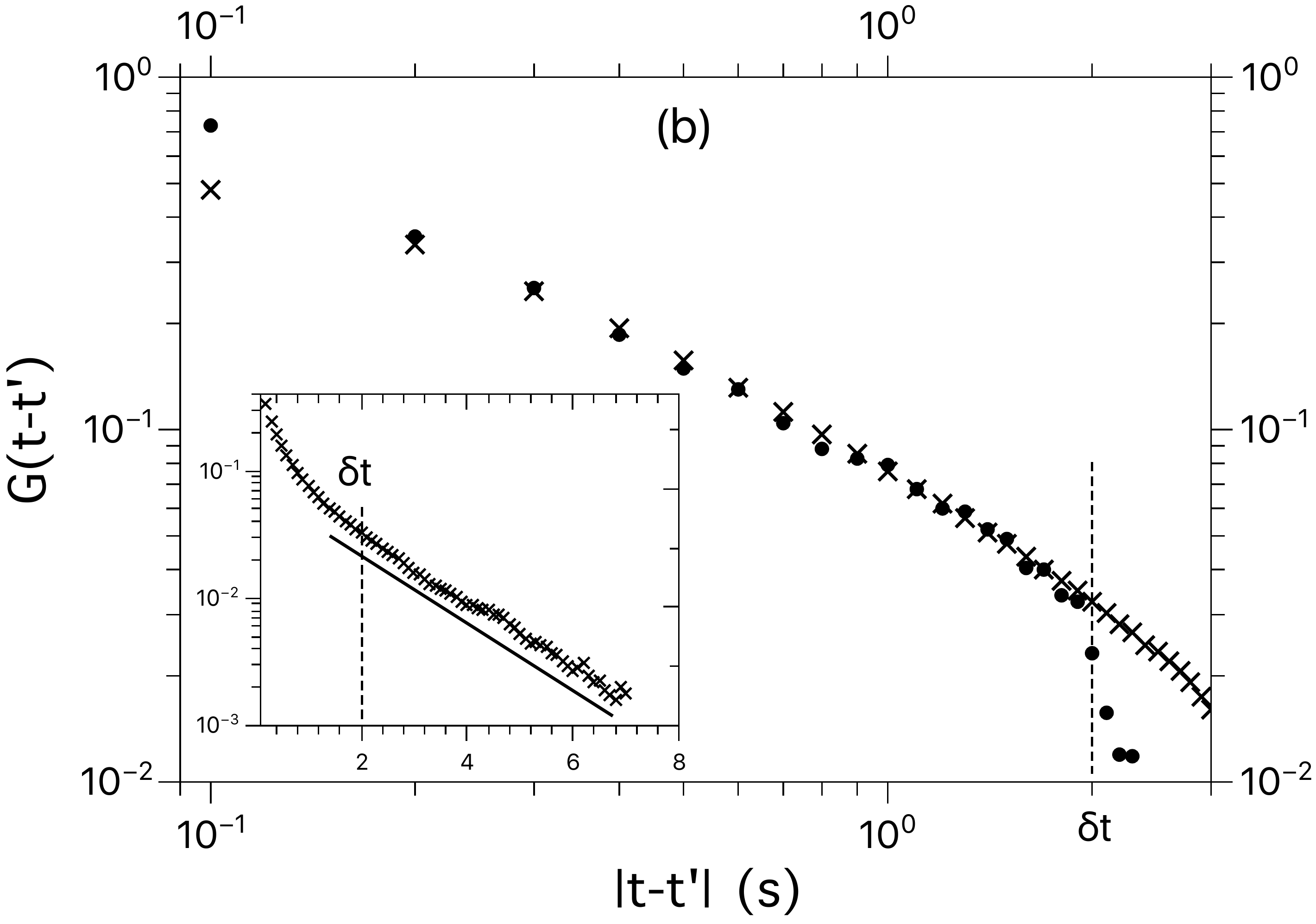}
\caption{Empirical (dots) and modeled (crosses) correlation functions $F(t-t')$ and $G(t-t')$. Crosses refer, in (a) and (b), respectively, to modeling parameters $p=0.86$ and {\hbox{$p=0.95$}}. The semi-log plot in the inset of (b) indicates the simple 
exponential form of $G(t-t')$ at large enough $|t-t'|$.}
\end{figure}

\noindent (i) We assume that we can model the observed coherence (time persistence) of low-speed streaks by a single mode-independent and not small probability parameter $p$, where {\hbox{$p=p_2=p_3=...=p_{10}$}};
\vspace{0.2cm}

\noindent (ii) $P_0$ turns out to be negligible, so we suppress transitions from the OS mode $m=1$ to $m=0$, by imposing that $p_1 = 1$ (other transitions to the mode $m=0$ from modes $m \neq 1$ are possible, but they are of $O((1-p)^2)$.
\vspace{0.2cm}

\noindent Therefore, we end up with 11 parameters ($q_0,q_1,..., q_9$ and $p$) to locate the minimum value of (\ref{qerror}). The result is a slightly overdetermined system, but if besides ${\mathbb{P}}_\infty$, the correlation functions $F(t-t')$ and $G(t-t')$ turn out to be well reproduced with the same set of probability parameters, as an extra bonus, then the model can be taken as physically appealing. That is the heuristic setup that we have in mind.

We have resorted to a straightforward Monte Carlo procedure to obtain the set of $q_m$'s that minimizes (\ref{qerror}) for various fixed values of $p$. We find, as shown in Fig.~5, that the quadratic error quickly drops for 
$p \geq 0.85$. The modeled asymptotic probabilities for the occurrence of OS modes are excellently compared, in Fig.~6, to the empirical ones for the cases $p=0.86$ and $p=0.95$. These are the values of $p$ that lead to good accounts of $F(t-t')$ and $G(t-t')$, as reported in Fig.~7. The related values of the probabilities $q_m$ are listed in Table II. Even if a point of subjective concern, the uncertainty of about 10$\%$ in the definition of $p$ should be taken as relatively small, {\it{vis a vis}} the model's accuracy in predicting the decaying profiles of the OS correlation functions.
\begin{table}[h]
\centering
  \begin{tabular}{ | c | c | c | c | c | c | c | c | c |  c | c |}
    \hline
    $p$ & $q_0$ & $q_1$ & $q_2$ &$q_3$ &$q_4$ &$q_5$ &$q_6$ &$q_7$ &$q_8$ &$q_9$ \\ \hline
    $0.86$ & $0.53$ & $0.96$ & $0.95$ &$0.92$ &$0.92$ &$0.85$ &$0.95$ &$0.75$ &$0.86$ &$1.0$ \\ \hline
        $0.95$ & $0.22$ & $0.98$ & $0.98$ &$0.97$ &$0.97$ &$0.96$ &$0.97$ &$0.93$ &$0.94$ &$0.49$ \\ 
    \hline
  \end{tabular}
\caption{The list of probabilities $q_m$'s which describe the persistence of inactive streak channels, for the cases {\hbox{$p=0.86$ and $p=0.95$}}.}
\end{table}

Also evidenced in the inset Fig.~7 is the exponential decay profile of the modeled $G(t-t')$ for time intervals larger than $\delta t$. At present, this point rests as a prediction of the modeling scenario introduced in this work, akin with the observed sudden undersampling of the time series for larger decimations. We note that the crossover to the faster exponential decay of correlation functions takes place at $\delta t \approx 2D/U$, where $U$ is the bulk flow velocity. Thus, the physical picture that emerges is that the OSs are packed as chains of low-speed streaks and vortical structures which are strongly correlated within sizes that scale with the pipe's diameter, although they are merged along the entire turbulent flow. 

To summarize, we have investigated the stochastic properties of the non-Markovian OS mode transitions in a turbulent pipe flow, recovering them as a surjective mapping of a lower-level Markov process. The essential idea that underlies the model construction is that a given OS mode may be associated to several spatial arrangements of its low-speed streaks into a fixed number of ``streak channels" which azimuthally partition the pipe's cross section. 

We find that the Markov model can account for the scaling behavior of specifically introduced correlation functions of OS mode transitions. Further work is in order, not only to enlarge the size of sPIV ensembles, but to address, in an analytical way, the very unexpected self-similar dynamics of the OS mode transitions. We point out that the dynamical scaling range of the recurrent OS transitions reflects the existence of finite-sized OS packets along the pipe flow, correlated at integral length scales (i.e., the pipe's diameter). 

An interesting theoretical direction to pursue is related to the use of instanton techniques \cite{grafke_etal} to evaluate the transition probabilities between unstable flow configurations as are the OS modes. In the turbulence or transitional context, instantons are taken, respectively, as extreme events or flow configurations that dominate the probability measures in the weak coupling limit. They have been successfully applied to a number of fluid dynamic problems, as in geophysical models, homogeneous turbulence and the laminar-turbulent transition in shear flows \cite{laurie_bouchet,apol_etal,gome_etal}. 

We conclude by noting that the findings here presented are likely to add relevant phenomenological information to the discussion of fundamentally important issues in pipe flow turbulence, as drag control and particle-laden dynamics, once they are closely connected to the statistical features of near-wall coherent structures \cite{choi_etal,schoppa_hussain,marusic_etal,gall_etal,wang_richter,brandt_coletti}.

%
%
%

\vspace{0.5cm}

\leftline{{\it{Acknowledgments}}}
\vspace{0.3cm}

This work was partially supported by the Conselho Nacional de Desenvolvimento Científico e Tecnológico (CNPq) and by Fundação Coppetec/UFRJ (project number 20459). L.M. thanks E. Marensi for enlightening discussions about the phenomenology of traveling waves.

\end{document}